\documentclass[a4paper,fleqn,usenatbib]{mnras}

\usepackage{newtxtext,newtxmath}

\usepackage[T1]{fontenc}
\usepackage{ae,aecompl}

\usepackage{graphicx}	
\usepackage{amsmath}	
\usepackage{amssymb}	
\usepackage{multirow}

\usepackage{bm}

\title[Gravitational lensing of GWs]{Gravitational lensing of gravitational waves: A statistical perspective}

\author[S. S. Li et al.]{
Shun-Sheng Li,$^{1,2}$\thanks{E-mail: lshuns@nao.cas.cn}
Shude Mao,$^{3,1,4}$
Yuetong Zhao,$^{1,2}$
and Youjun Lu$^{1,2}$
\\
$^{1}$National Astronomical Observatories, Chinese Academy of Sciences, Beijing 100012, China\\
$^{2}$School of Astronomy and Space Science, University of Chinese Academy of Sciences, Beijing 100049, China\\
$^{3}$Physics Department and Tsinghua Centre for Astrophysics, Tsinghua University, Beijing 100084, China\\
$^{4}$Jodrell Bank Centre for Astrophysics, School of Physics and Astronomy, The University of Manchester, Oxford Road, Manchester M13 9PL, UK
}

\date{Accepted 2018 February 14. Received 2018 February 14; in original form 2018 January 3}

\pubyear{2018}

\begin{document}
\label{firstpage}
\pagerange{\pageref{firstpage}--\pageref{lastpage}}
\maketitle

\begin{abstract}
In this paper, we study the strong gravitational lensing of gravitational waves (GWs) from a statistical perspective, with particular focus on the high frequency GWs from stellar binary black hole coalescences. These are most promising targets for ground-based detectors such as Advanced Laser Interferometer Gravitational Wave Observatory (aLIGO) and the proposed Einstein Telescope (ET) and can be safely treated under the geometrical optics limit for GW propagation. We perform a thorough calculation of the lensing rate, by taking account of effects caused by the ellipticity of lensing galaxies, lens environments, and magnification bias. We find that in certain GW source rate scenarios, we should be able to observe strongly lensed GW events once per year ($\sim1~\text{yr}^{-1}$) in the aLIGO survey at its design sensitivity; for the proposed ET survey, the rate could be as high as $\sim80~\text{yr}^{-1}$. These results depend on the estimate of GW source abundance, and hence can be correspondingly modified with an improvement in our understanding of the merger rate of stellar binary black holes. We also compute the fraction of four-image lens systems in each survey, predicting it to be $\sim30$ per cent for the aLIGO survey and $\sim6$ per cent for the ET survey. Finally, we evaluate the possibility of missing some images due to the finite survey duration, by presenting the probability distribution of lensing time delays. We predict that this selection bias will be insignificant in future GW surveys, as most of the lens systems ($\sim90$ per cent) will have time delays less than $\sim1$ month, which will be far shorter than survey durations.
\end{abstract}




\section{Introduction}
The four signals of gravitational waves (GWs) from binary black hole systems, GW150914~\citep{Abbott2016}, GW151226~\citep{Abbott2016b} , GW170104~\citep{Abbott2017}, and GW170608~\citep{Abbott2017d} detected by Advanced Laser Interferometer Gravitational Wave Observatory (aLIGO) during its first and second observing runs (O1, O2), marked the commencement of GW astronomy. More recently, with the Advanced Virgo detector becoming operational, we had the first joint detection GW170814~\citep{Abbott2017b} and the first binary neutron star (BNS) signal GW170817~\citep{Abbott2017c}. These observations provide us a new opportunity to study astrophysics and cosmology.

Since \citet{Wang1996} proposed the possibility of observing several strongly lensed GW events in the context of aLIGO type detectors, gravitational lensing of GWs has been widely discussed over the past two decades. Such discussions involve diffraction effects in lensed GW events \citep{Nakamura1998,Takahashi2003}, the waveform distortion caused by the gravitational lensing \citep{Cao2014,Dai2017}, the influence on the statistical signatures of black hole mergers \citep{Dai2017a} as well as the potential for studying fundamental physics \citep{Collett2017,Fan2017} and cosmology \citep{Sereno2011,Liao2017,Wei2017}. Nevertheless, in spite of the broad range of topics discussed so far, the field of gravitational lensing of GWs is still worth an extensive exploration in order to fully understand the phenomenon and how to employ it to investigate the Universe.

One crucial question we have to answer before a further exploration of gravitational lensing of GW occurs is `how many lensed GW events are expected to be observed?' Indeed, several discussions on this aspect already exist in the literature. For example, \citet{Sereno2010} studied lensed GW events from the merging of massive black hole binaries in the context of the LISA mission; \citet{Biesiada2014} considered the observational context for the \emph{Einstein Telescope} (ET); and more recently, \citet{Ng2017} revisited the LIGO lensing rate. However, all the studies mentioned so far adopt the simplest lens model, which treats the lens mass distribution as axisymmetric.

In this paper, we present some extensions to the calculation of the lensed GW rate, making allowances for the ellipticity of the lens, the lens environment (as an external shear), and for magnification bias. This treatment not only provides a more precise prediction about the lensing rate, including more statistical properties, but also can serve as a useful tool for cosmological study (e.g. \citealt{Chae2003}). We concentrate on the ground-based GW detectors, specifically aLIGO and the proposed ET. Nevertheless, the strategy developed here is general and can be easily extended to address other similar GW surveys as long as the geometrical optics approximation to GW propagation is valid.

The estimate of source rate dominates the prediction for the lensed event rate. Here we consider GWs from the coalescence of stellar binary black holes as the only sources, since they are the main signals received by ground-based detectors \citep{Dominik2013, Abbott2016c}. In order to obtain the source rate, we use a similar approach as in \citet{CaoL2017} to estimate the merger rate of stellar binary black holes, and then use the GW detection theory developed by \citet{Finn1996} to translate the intrinsic merger rate into the detectable source rate.

Another essential factor that can affect the observation of lensed events is the lensing time delay, as an image with time delay comparable to the survey's duration has a high probability of being missed by the detector. We assess this selection bias by computing the distribution function of time delays corresponding to the lens properties adopted in this paper.

Our paper is organized as follows. In Section~\ref{sec:model}, we describe the approach to lensing rate calculation and the assumption of lens properties. We present our results in Section~\ref{sec:results} and summary in Section~\ref{sec:summary}. Throughout this paper, we adopt geometric units with $G=c=1$ and assume a Lambda cold dark matter universe with $(\Omega_M,\Omega_{\Lambda})=(0.3,0.7)$ and a Hubble parameter $H_0=70~\text{km}~ \text{s}^{-1}~\text{Mpc}^{-1}$.

\section{Theoretical Model}
\label{sec:model}

In this section, we present our lens model (Section~\ref{lensM}) and our GW detection model (Section~\ref{GWM}). With the theory of lensing statistics (Section~\ref{lenss}), we then derive the formulae to calculate the expected lensing rate in Section~\ref{Expec}. The theory developed in this section is general and can be used to estimate strong gravitational lensing rates in any ground-based GW surveys so long as the geometrical optics approximation (see below) is valid.

\subsection{Lens modelling}
\label{lensM}
When the lens mass is larger than $\sim 10^5M_{\odot}(f/Hz)^{-1}$ where $f$ is the frequency of the incident waves, the propagation of GWs is analogous to that of light. This is known as the geometrical optics approximation to GW propagation \citep{Takahashi2003}. Since we here concentrate on the macrolensing by galaxies ($M\gtrsim 10^{10}M_{\odot}$) of high frequency GWs ($f\gtrsim 10$Hz), this condition is always satisfied. Hence, it is a reasonable approximation in the context of this paper to neglect the wave effect and adopt the standard optical gravitational lens theory to study the gravitational lensing of GWs.

As it is broadly reckoned that the strong lensing probability is dominated by early-type galaxies (\citealt{Turner1984}; \citealt{Moller2007}, and references therein), we only consider early-type galaxies as lensing objects. The singular isothermal ellipsoid (SIE) is adopted to model the mass distributions of the lensing galaxies. For the SIE convergence in Cartesian coordinates ($x,y$), we adopt the form developed by \citet{Keeton1998}:
\begin{equation}\label{SIE}
\kappa(x,y) = \frac{1}{2}\frac{\lambda(q)\sqrt{q}}{\sqrt{x^2+q^2y^2}}~,
\end{equation}
where $q$ is the projected minor-to-major axis ratio, and $\lambda(q)$, the so-called `dynamical normalization', depends on the three-dimensional shape of lensing galaxies \citep{Chae2003}.

Furthermore, we consider the influence from the lens environment as an external shear $\bm{\gamma}$ whose lens potential is given by (\citealt{Kochanek1991}; \citealt{Witt1997}, and references therein)
\begin{equation}\label{shear}
\begin{split}
\phi^{\rm shear}=&\frac{\gamma}{2}(x^2-y^2)\cos 2\theta_{\bm{\gamma}}+\gamma xy \sin 2\theta_{\bm{\gamma}}\\
=&\frac{1}{2}(x^2-y^2)\gamma_1+xy\gamma_2~,
\end{split}
\end{equation}
where $(\gamma_1,\gamma_2)$ are the two components of the shear in Cartesian coordinates, and $(\gamma,\theta_{\bm{\gamma}})$ are the corresponding amplitude and direction components in polar coordinates. The connections between these two coordinate systems are: $\gamma_1 =\gamma\cos 2\theta_{\bm{\gamma}} ~,~ \gamma_2=\gamma\sin 2\theta_{\bm{\gamma}}$.

More detailed discussions of the lens model can be found in Appendix \ref{appGL}.

\subsection{GW modelling}
\label{GWM}
An estimate of the GW event rate density is required for calculating the expected number of lensed events. This involves the theory of GW detection, which has been discussed by many authors \citep{Finn1993,Finn1996,Flanagan1998,Taylor2012}. Here we mainly follow the framework developed by \citet{Finn1996}.

For Gaussian and stationary noise, the optimal matched filtering signal-to-noise ratio (S/N) $\rho$ is defined as (e.g. \citealt{Flanagan1998})
\begin{equation}\label{SNRdef}
\rho^2 = 4\int_{0}^{\infty}df~\frac{|h(f)|^2}{S_n(f)}~,
\end{equation}
where $S_n(f)$ is the one-sided power spectral density of the detector's noise, and $h(f)$ is the Fourier transform of the detector's response to the GWs.

The GW generated by an inspiralling binary system can be approximately described by a quadrupolar formula (Newtonian order) with the frequency twice the binary's orbital frequency. This waveform model does not meet the empirical requirement coming from the analysis of GW data, but is accurate enough for our statistical purpose. The amplitude given by the quadrupolar formula can be written as \citep{Taylor2012}
\begin{equation}\label{waveform}
|h(f)|=\frac{1}{D_L}\left(\frac{5}{24}\right)^{1/2}\left(\frac{\mathcal{M}_z^{5}}{{{\rm\pi}}^{4}}\right)^{1/6}\left(\frac{\Theta}{4}\right)f^{-7/6}~,
\end{equation}
where $D_L$ is the luminosity distance and
\begin{equation}
\mathcal{M}_z \equiv (1+z)\mathcal{M}_0= (1+z)\frac{(m_1m_2)^{3/5}}{(m_1+m_2)^{1/5}}
\end{equation}
is the observed (redshifted) chirp mass with $\mathcal{M}_0$ the intrinsic chirp mass, and $\Theta$ is the orientation function:
\begin{equation}
\Theta \equiv 2[F_{+}^2(1+\cos^2i)^2+4F_{\times}^2\cos^2i]^{1/2}~,
\end{equation}
with
\begin{equation}
\begin{split}
F_{+}\equiv &\frac{1}{2}(1+\cos^2\theta)\cos 2\phi\cos 2 \psi - \cos\theta\sin 2\phi \sin 2\psi~,\\
F_{\times}\equiv &\frac{1}{2}(1+\cos ^2\theta)\cos 2\phi\sin 2 \psi + \cos\theta\sin 2 \phi \cos 2\psi
\end{split}
\end{equation}
describing the detector's responses to the different GW polarizations. Obviously, $\Theta$ depends only on the sky position and relative orientation of the source to the detector ($\theta, \phi, i, \psi$)\footnote{$\theta$ and $\phi$ correspond to the usual spherical coordinates that describe the direction to the source, while $i$ and $\psi$ give the source's orientation with respect to the detector (see \citealt[Section \uppercase\expandafter{\romannumeral2}-C]{Finn1996} for a detailed discussion).}, which are uncorrelated and uniformly distributed. A reasonable approximation of the probability distribution of $\Theta$ is given by \citep{Finn1996}
\begin{equation}\label{Thetadis}
P_{\Theta}(\Theta)=\left\{ \begin{array}{ll}
5\Theta(4-\Theta)^3/256 & \text{if } 0<\Theta<4 \\
0 & \text{otherwise}~.\\
\end{array}\right.
\end{equation}

Combining equation (\ref{SNRdef}) with equation (\ref{waveform}), the S/N can be written as \citep{Finn1996,Taylor2012}
\begin{equation}\label{SNR2}
\rho = 8\Theta\frac{R_0}{D_L}\left(\frac{\mathcal{M}_z}{1.2M_{\odot}}\right)^{5/6}\sqrt{\zeta(f_{\rm max})}~,
\end{equation}
where
\begin{equation}\label{R0}
R_0^2\equiv \frac{5}{192{\rm\pi}}\left(\frac{3}{20}\right)^{5/3}x_{7/3}M_{\odot}^2
\end{equation}
is the detector's characteristic distance parameter, with
\begin{equation}
x_{7/3}\equiv \int_{0}^{\infty}\frac{({\rm\pi} M_{\odot})^2}{({\rm\pi} f M_{\odot})^{7/3}S_n(f)}df~,
\end{equation}
and
\begin{equation}
\label{equ:zeta}
\zeta(f_{\rm max})\equiv \frac{1}{x_{7/3}}\int_{0}^{2f_{\rm max}}df~\frac{({\rm\pi} M_{\odot})^2}{({\rm\pi} f M_{\odot})^{7/3}S_n(f)}
\end{equation}
is the dimensionless function reflecting the overlap between the GW signal generated by the inspiral stage and the detector's effective bandwidth: $\zeta(f_{\rm max})$ is unity if $2f_{\rm max}$ is larger than the upper bound frequency of the detector's bandwidth (i.e., the GW signal from the inspiral stage completely covers the detector's effective bandwidth), and is less than unity if the inspiral terminates within the detector's bandwidth.

The argument $f_{\rm max}$ is the redshifted orbital frequency at which the quadrupolar formula is no longer applicable (the binary finishes the inspiral and starts to merge). It is plausible to choose the entering of the innermost circular orbital (ICO) as the end of the inspiral stage. For binaries with equal-mass, this can be described as \citep{Taylor2012}
\begin{equation}\label{fmax}
f_{\rm max}=\frac{f_{\text{ICO}}}{1+z}=\frac{785~\text{Hz}}{1+z}\left(\frac{2.8M_{\odot}}{M}\right)~,
\end{equation}
where $M$ is the total mass of the binary. For binaries with unequal mass, $f_{\text{ICO}}$ also depends on the mass asymmetry. In our simulation, we ignore this small correction, and make exclusive use of equation (\ref{fmax}). We calculate $\zeta(f_{\rm max})$ for the typical total mass in our source sample ($M=10M_{\odot}$) and find it to be close to unity ($\sim 0.98$). Hence for simplicity, we adopt $\zeta(f_{\rm max})=1$ in the following calculations.

The distribution of the GW event rate in the observer's frame with $z,\mathcal{M}_0$ and $\rho$ is given by \citep{Finn1996}
\begin{equation}\label{ddrapp}
\frac{d^3\dot{N}}{dzd\mathcal{M}_0d\rho}=\frac{dV^c}{dz}\frac{R_{\rm mrg}(\mathcal{M}_0;z)}{(1+z)}P_{\rho}(\rho|z,\mathcal{M}_0)~,
\end{equation}
where $dV^c$ is the differential comoving volume and the factor $1/(1+z)$ accounts for the time dilation. $R_{\rm mrg}(\mathcal{M}_0;z)$ is the intrinsic merger rate density with respect to the chirp mass $\mathcal{M}_0$ at redshift $z$. Our model to estimate this density is presented in Appendix~\ref{source}. The distribution $P_{\rho}(\rho|z,\mathcal{M}_0)$ can be calculated by combining equations (\ref{Thetadis}) and (\ref{SNR2}):
\begin{equation}\label{disrho}
\begin{split}
P_{\rho}(\rho|z,\mathcal{M}_0) =& P_{\Theta}(\Theta)\frac{\partial \Theta}{\partial \rho}{\bigg|}_{\mathcal{M}_0,z}\\
=&P_{\Theta}[\Theta_{\rho}]\frac{\Theta_{\rho}}{\rho}~,
\end{split}
\end{equation}
where $\Theta_{\rho}$ is rearranged from equation (\ref{SNR2}):
\begin{equation}
\Theta_{\rho}= \frac{\rho}{8}\frac{D_L}{R_0}\left(\frac{1.2M_{\odot}}{\mathcal{M}_z}\right)^{5/6}\frac{1}{\sqrt{\zeta(f_{\rm max})}}~.
\end{equation}

By marginalizing over $\mathcal{M}_0$ in equation (\ref{ddrapp}), we can obtain the differential GW event rate with S/N $\rho$ at redshift $z$:
\begin{equation}
\label{GWdis}
\Phi(\rho;z)=\int d\mathcal{M}_0~ \frac{d^3\dot{N}}{dzd\mathcal{M}_0d\rho}~.
\end{equation}

The GW event rate for a particular detector of threshold $\rho_0$ is given by
\begin{equation}\label{GWrate}
  \dot{N}_{s}(>\rho_0)=\int_{0}^{\infty}dz_s\int_{\rho_0}^{\infty}d\rho ~ \Phi(\rho;z_s)~,
\end{equation}
and the corresponding differential rate is
\begin{equation}\label{GWrated}
  \frac{d\dot{N}_{s}(>\rho_0)}{dz}=\int_{\rho_0}^{\infty}d\rho ~ \Phi(z_s,\rho)~.
\end{equation}

\subsection{Lensing statistics}
\label{lenss}
In the context of lensing statistics, the most important parameter is the so-called optical depth, or the differential lensing probability (e.g. \citealt{Turner1984, Chae2003,Huterer2005}):
\begin{equation}
\label{equa:opticaldepth}
\begin{split}
p(\rho;z_s) = &\frac{1}{4{\rm\pi}}\int_0^{z_s}dV^c\int_0^{\infty} d\sigma_v~ \Psi(\sigma_v)\\
&\times \int  dq~p_q(q)\iint d\bm{\gamma} ~p_{\bm{\gamma}}(\gamma,\theta_{\bm{\gamma}})\\
&\times B(\rho;z_s)~\sigma_{\ell}(\sigma_v,z_{\ell},z_s,\bm{\gamma},q)~,
\end{split}
\end{equation}
which describes the differential probability for a given source with S/N $\rho$ at redshift $z_s$ to be lensed.

The first integral takes into account the comoving volume between the observer and the source. It is required for calculating the total number of lensing galaxies.

The second integral gives the number density of lensing galaxies in comoving volume, where $\Psi(\sigma_v)$ is the velocity distribution function of lensing galaxies. In the context of lensing statistics, the modified Schechter function \citep{Choi2007}
\begin{equation}
\label{equ:VDF}
\Psi(\sigma_v)=\phi_{\ast}\left( \frac{\sigma_v}{\sigma_{\ast}} \right)^{\alpha}\exp \left[-\left(\frac{\sigma_v}{\sigma_{\ast}}\right)^{\beta}\right]\frac{\beta}{\Gamma(\alpha/\beta)}\frac{1}{\sigma_v}
\end{equation}
is often used to fit the velocity distribution function, where $(\phi_{\ast}, \sigma_{\ast}, \alpha, \beta)$=$(8.0\times 10^{-3}h^3~\text{Mpc}^{-3}, 161\text{ km s}^{-1}, 2.32, 2.67)$.

The third integral is over the distribution $p_q(q)$ of the projected axis ratio $q$. We adopt a Gaussian distribution to describe $p_q(q)$, with a mean of $0.7$, and standard deviation of $0.16$. The distribution is truncated at $q=0.2$ and $1.0$ . This is consistent with the observations \citep{Jorgensen1995,Sheth2003}.

The fourth integral is two-dimensional, where $p_{\bm{\gamma}}(\gamma,\theta_{\bm{\gamma}})$ denotes the distribution of external shear $\bm{\gamma}$. Following \citet{Huterer2005}, we assume the amplitude $\gamma$ follows a log-normal distribution with mean $\ln 0.05$ and standard deviation $0.2$ (note: the mean and standard deviation are not the values for $\gamma$ itself, but of the underlying normal distribution it is derived from). The direction $\theta_{\bm{\gamma}}$ is assumed to be random.

The bias factor $B(\rho;z_s)$ describes an enhancement of the representation of events due to the magnification caused by the lens (magnification bias):\footnote{In gravitational lensing of GWs, the amplification in S/N is $\sqrt{\mu}$, since we directly observe the waveform instead of intensity.}
\begin{equation}
\label{bias}
\begin{split}
B(\rho;z_s)=\frac{\Phi_ed\rho}{\Phi_dd\rho}&=\frac{\int_0^{\infty}d\mu~p_{\mu}(\mu)~\Phi(\rho/\sqrt{\mu};z_s)d\rho/\sqrt{\mu}}{\Phi(\rho;z_s)d\rho}\\
&=\int_0^{\infty}d\mu~\frac{p_{\mu}(\mu)}{\sqrt{\mu}}\frac{\Phi(\rho/\sqrt{\mu};z_s)}{\Phi(\rho;z_s)}~.
\end{split}
\end{equation}
Note that $\int_0^{\infty}d\mu~p_{\mu}(\mu)=1$ is required, in order to combine the bias factor into optical depth naturally.

The choice of the magnification factor $\mu$ is demanded for multiple-image systems in calculation of the bias factor. For double (two-image) lenses, we adopt the magnification factor of the fainter image as $\mu$, so that both images are magnified above the threshold. For quadruple (four-image) lenses, we adopt the magnification factor of the third brightest image, hence at least three images are magnified above the threshold. Also, this choice ensures the detection of the first lensed image to arrive, since the third brightest image is generally expected to arrive first \citep{Oguri2010}.

We calculate the cross-section $\sigma_{\ell}$ in angular dimensions, that results in the normalization factor $1/4{\rm\pi}$ in equation (\ref{equa:opticaldepth}). It is convenient to define a dimensionless cross-section $\hat{\sigma}_{\ell}\equiv \sigma_{\ell}/\theta_E^2$, where $\theta_E \equiv 4{\rm\pi}\sigma_v^2(D_{\ell s}/D_{s})$ is the angular Einstein radius with $D_s$, $D_{\ell s}$ denoting the angular distances to the source and between the lens and the source, respectively. Due to the feature of isothermal lens models that the dependence on $z_s$, $z_{\ell}$, $\sigma_v$ is contained in $\theta_E$, the dimensionless cross-section $\hat{\sigma}_{\ell}$ depends only on $(\bm{\gamma},q)$.
And the optical depth now can be rearranged to a more practical form:
\begin{equation}\label{opt2}
\begin{split}
p(\rho;z_s) = &\frac{1}{4{\rm\pi}}\left[\int_0^{z_s}dV^c\int_0^{\infty} d\sigma_v~ \Psi(\sigma_v) ~\theta_E^2(\sigma_v,z_{\ell},z_s)\right]\\
&\times\left[ \int dq~ p_q(q) \iint d\bm{\gamma} ~p_{\bm{\gamma}}(\gamma,\theta_{\bm{\gamma}})~B(\rho;z_s)~
\hat{\sigma}_{\ell}(\bm{\gamma},q)\right]~.
\end{split}
\end{equation}
The two parts separated by square brackets are independent, and can be integrated separately.

As for the problem of determining the region of cross-section, we handle double and quadruple lenses separately.\footnote{The naked cusp lenses are ignored in this paper, since they seldom happen at galaxy-scale lenses \citep{Oguri2010}.} This treatment gives us the fraction of quadruple lenses. For double lenses, the condition that the fainter image is magnified above threshold $\rho_0$ is taken to define the region of cross-section. This treatment guarantees the theoretical detectability of multiple images.

\subsection{Expected lensing rate}
\label{Expec}
The expected lensing rate for a particular detector of threshold $\rho_0$ can be calculated as
\begin{equation}\label{event1}
\dot{N}_{\ell}(>\rho_0)=\int_{0}^{\infty}dz_s\int_{\rho_0}^{\infty}d\rho ~ p(\rho;z_s)~\Phi(\rho;z_s).
\end{equation}

In theory, by substituting equations (\ref{GWdis}) and (\ref{opt2}) into equation (\ref{event1}), we can obtain the lensing rate as a function of threshold $\rho_0$. In practice, it is numerically more friendly if some rearrangements or reductions are undertaken. Hence we introduce a more practical form for calculation of the expected lensing rate:
\begin{equation}\label{event2}
\begin{split}
\dot{N}_{\ell}(>\rho_0) =&\int_{0}^{\infty} dz_s\int_{\rho_0}^{\infty}d\rho\int dq~ p_q(q) \iint d\bm{\gamma} ~p_{\bm{\gamma}}(\gamma,\theta_{\bm{\gamma}})\\
&\times\iint~d\bm{u}~f(\rho,q,\bm{\gamma},\bm{u};z_s)~,
\end{split}
\end{equation}
and the last integral which combines the dimensionless cross-section $\hat{\sigma}_{\ell}$, the bias factor $B(\rho;z_s)$, and the differential GW event rate $\Phi(\rho;z_s)$ is performed over the determined cross-section region $\mathbf{u}$.  While
\begin{equation}
 f(\rho,q,\bm{\gamma},\bm{u};z_s)=\frac{g(z_s)}{\sqrt{\mu}}\Phi(\rho/\sqrt{\mu};z_s)~,
\end{equation}
where
\begin{equation}
g(z_s)=\frac{1}{4{\rm\pi}}\left[\int_0^{z_s}dV^c\int_0^{\infty} d\sigma_v~ \Psi(\sigma_v) ~\theta_E^2(\sigma_v,z_{\ell},z_s)\right]
\end{equation}
is the integral which combines the velocity distribution function $\Psi(\sigma_v)$ and the angular Einstein radius $\theta_{E}$.

For the differential rate, it is convenient to write the function as
\begin{equation}\label{eventd}
\frac{d\dot{N}_{\ell}(>\rho_0)}{dz_s} = g(z_s)~h(>\rho_0;z_s)~,
\end{equation}
where
\begin{equation}
\begin{split}
h(>\rho_0;z_s)=&\int_{\rho_0}^{\infty}d\rho\int dq~ p_q(q) \iint d\bm{\gamma} ~p_{\bm{\gamma}}(\gamma_1,\gamma_2)\\
&\times\iint\frac{d\mathbf{u}}{\sqrt{\mu}}\Phi(\rho/\sqrt{\mu};z_s)~.
\end{split}
\end{equation}

\section{Results}
\label{sec:results}

In this section, we present our prediction of the strongly lensed GW event rate (Section~\ref{res:event}). We take aLIGO operating at its design sensitivity and the ET utilising its `xylophone' configuration as illustrations. We calculate the lensing rate as a function of the characteristic distance $R_0$ (see below) to give a more general prediction for ground-based detectors with arbitrary sensitivity. In Section~\ref{res:time}, we illustrate the probability distribution of lensing time delays to assess the detectability of multiple images in a finite duration.

\subsection{Event rate}
\label{res:event}

The lensing rate is strongly dependent on the estimate of the GW event rate, which, in turn, depends on the estimate of the merger rate of stellar binary black holes. As an illustration, we use a simple recipe analogous to that in \citet{CaoL2017} to compute the merger rate (see Appendix~\ref{source} for further details). Fig.~\ref{fig:intrinsic} shows our results on the merger rate density distribution as a function of cosmic time (redshift). The two different lines represent estimates obtained by using observationally determined star formation rate (SFR) functions from \citet{Strolger2004} (red solid line) and from \citet{Madau2014} (black dashed line). The discrepancy between these two results is noticeable at high redshift. This contradiction accounts for all the disparities in the following results. Despite the simplicity of our model, our results are comparable to those estimated through more sophisticated population synthesis models (see the comparison in \citealt{CaoL2017}).

The GW detector's sensitivity is described by a characteristic distance $R_0$, which depends only on the detector's noise power spectral density $S_n(f)$ (see equation~\ref{R0}). Generally speaking, the larger $R_0$, the farther a detector can observe. For aLIGO, we use the data `ZERO{\_}DET{\_}high{\_}P.txt' from \citet{Shoemaker2010} as the $S_n(f)$ for each interferometer operating at the design sensitivity. Since aLIGO consists of two interferometers with equal configurations and closely parallel orientations (one at Hanford, WA, and the other at Livingston, LA), we can treat aLIGO as a whole with the characteristic distance $\sqrt{2}$ times larger than that of each signal interferometer \citep{Finn1996}. For the ET, which uses 3rd-generation technology and the `xylophone' configuration, we adopt $R_0=1591$~Mpc \citep{Taylor2012}.

Once the intrinsic merger rate of stellar binary black holes and the characteristic distance of the detector are determined, we can obtain the unlensed and lensed GW event rates through equations (\ref{GWrate}) and (\ref{event2}), respectively. The threshold $\rho_0$ is set to be eight, that means a signal is identified as detected when its S/N is above eight. Table~\ref{tab:lensed} summarizes our results for the unlensed and lensed GW event rates in various detectors.

We predict that the unlensed GW event rates are $\sim 10^3\text{ yr}^{-1}$ for aLIGO at its design sensitivity ($R_0=155.4$~Mpc) and $\sim 10^5\text{ yr}^{-1}$ for ET ($R_0=1591$~Mpc). For comparison, we use the same strategy to compute the GW event rate at aLIGO's O2 run ($R_0=63.7$~Mpc, $\rho_0=13$, $\zeta= 0.4\sim 1.0$)\footnote{The threshold $\rho_0$ is set according to GW170104, which has the lowest S/N among aLIGO's O2 detections. The lower limit of $\zeta$ factor (see equation \ref{equ:zeta}) is calculated using the total mass of GW170814 which has the largest mass among aLIGO's O2 detections.} and obtain $15\sim 75\text{ yr}^{-1}$, which is consistent with the current aLIGO detection rate. The improved sensitivity and the lower threshold of S/N account for the much higher expected source rate at aLIGO's design sensitivity compared with that of the O2 run.

Based on these estimates of the GW event rate, we find that gravitational lensing of GWs is promising for both aLIGO at its design sensitivity and the proposed ET. More specifically, when the SFR function from \citet{Strolger2004} is adopted, both detectors have the largest expected numbers of lensed events (aLIGO $\sim 1\text{ yr}^{-1}$ and ET $\sim 80\text{ yr}^{-1}$). For the SFR function adopted from \citet{Madau2014}, the number in ET declines dramatically to $\sim 40\text{ yr}^{-1}$ due to the lower source rate expected at high redshift. The number in aLIGO drops only slightly and is still close to  $1\text{ yr}^{-1}$, since aLIGO is insensitive to the event rate at high redshift. Furthermore, we compute the fraction of quadruple lenses in each survey. Our calculation indicates that the quadruple fraction is approximately $30$ per cent for aLIGO events and $6$ per cent for ET events. The higher quadruple fraction in aLIGO corresponds to the larger magnification bias.

\begin{figure}
	\includegraphics[width=\columnwidth]{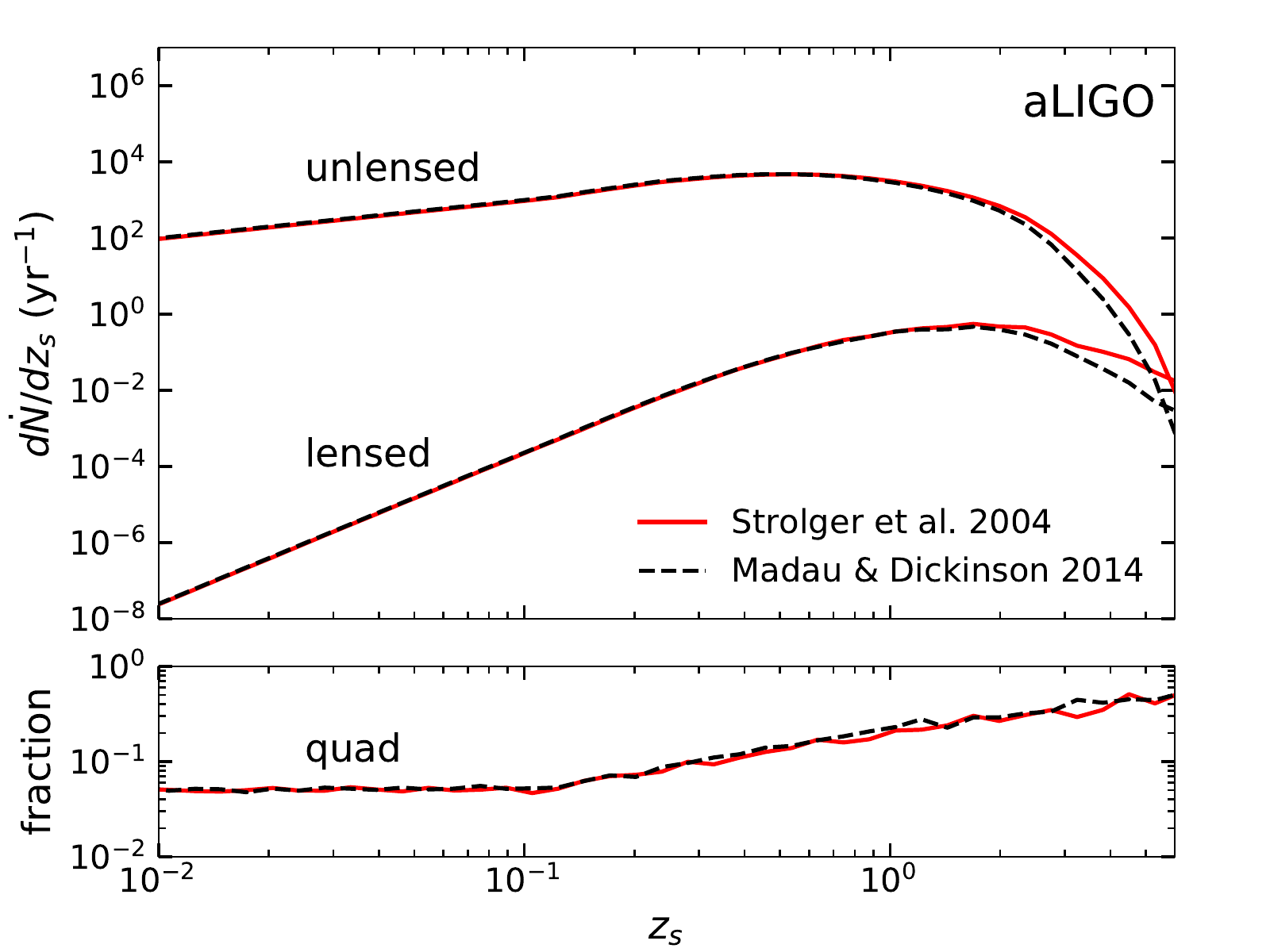}
    \caption{The differential rate of unlensed and lensed GW events (top panel) and the fraction of quadruple lenses (bottom panel) as a function of source redshift for aLIGO ($R_0=155.4$~Mpc). The red solid and black dashed lines represent results obtained by using the SFR function from \citet{Strolger2004} and from \citet{Madau2014}, respectively. The threshold of S/N is set to be eight.}
    \label{fig:lensed_aLIGO}
\end{figure}

\begin{figure}
	\includegraphics[width=\columnwidth]{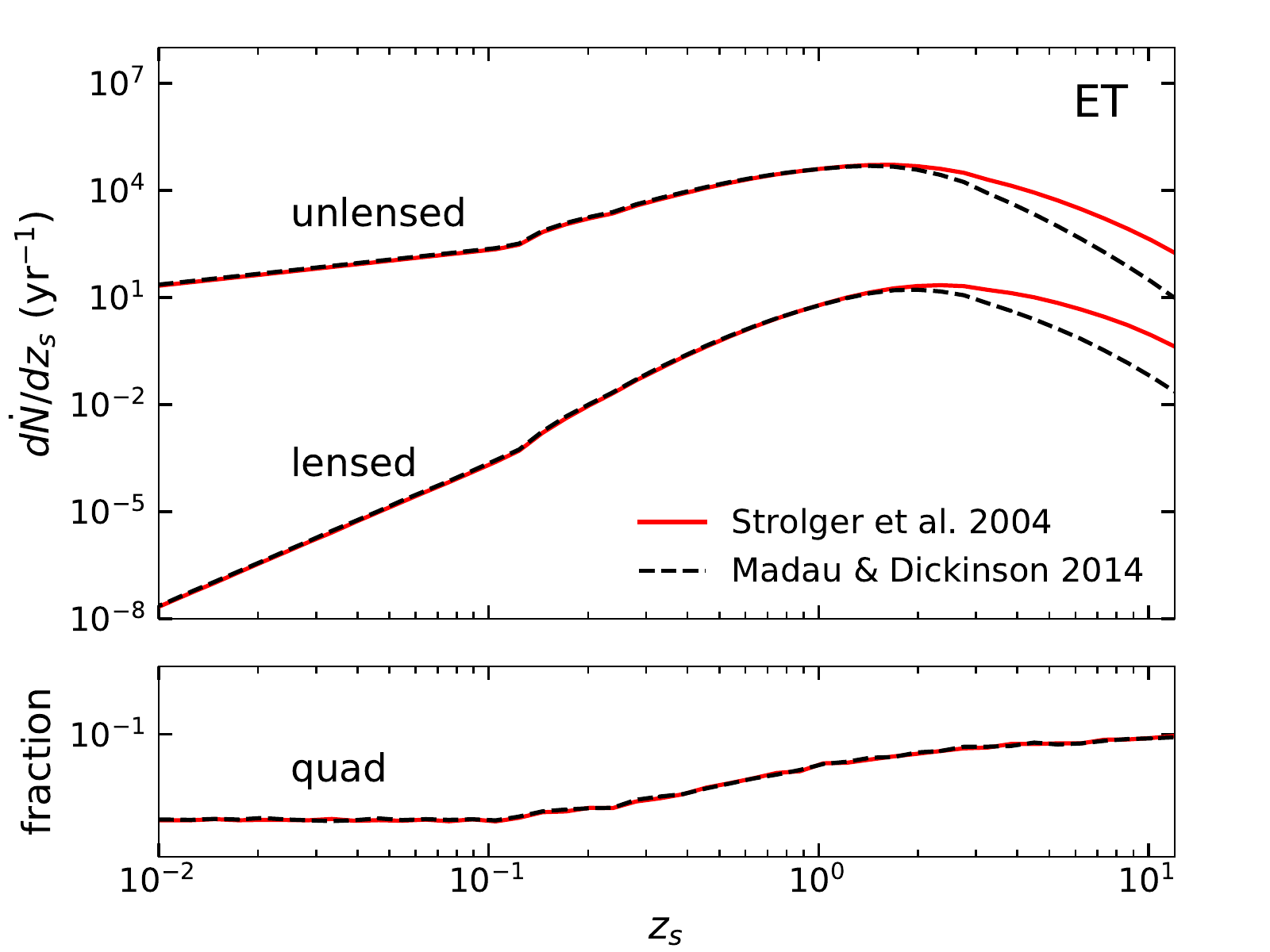}
    \caption{The differential rate of unlensed and lensed GW events (top panel) and the fraction of quadruple lenses (bottom panel) as a function of source redshift for ET ($R_0=1591$~Mpc). The red solid and black dashed lines represent results obtained by using the SFR function from \citet{Strolger2004} and from \citet{Madau2014}, respectively. The threshold of S/N is set to be eight.}
    \label{fig:lensed_ET}
\end{figure}

In Fig.~\ref{fig:lensed_aLIGO} (top panel), we plot the differential rates of unlensed and lensed GW events as a function of source redshift for aLIGO detector. The two different lines represent the results calculated by adopting the SFR functions from \citet{Strolger2004} (red solid line) and from \citet{Madau2014} (black dashed line). Roughly speaking, when the magnification bias is negligible, we have a scaling relationship between the differential rates of unlensed and lensed events, $d\dot{N}_{\ell}/dz_s\propto z_s^3 \cdot d\dot{N}_{s}/dz_s$, since the optical depth satisfies $p(z_s)\propto z_s^3$. This scaling roughly matches the slope in Fig.~\ref{fig:lensed_aLIGO} at low redshift. At high redshift, the trend of the lensed events is dominated by the magnification bias. In the bottom panel, we show the fraction of quadruple lenses as a function of source redshift. The rising quadruple fraction results from the increase in the magnification bias. Fig.~\ref{fig:lensed_ET} is the same as Fig.~\ref{fig:lensed_aLIGO} but for ET detector.

We also calculate the most probable redshifts of the lensed sources (from $16$ per cent to $84$ per cent) and find it ranges from $\sim1.1$ to $\sim2.7$ for aLIGO events and from $\sim1.5$ to $\sim3.7$ for ET events based on the SFR function from \citet{Madau2014}. If the SFR function from \citet{Strolger2004} is adopted, the redshifts are slightly higher due to the higher estimates of source rates at high redshift (see Fig.~\ref{fig:intrinsic}), ranging from $\sim1.2$ to $\sim3.3$ for aLIGO events and from $\sim1.8$ to $\sim5.7$ for ET events.

\begin{figure}
	\includegraphics[width=\columnwidth]{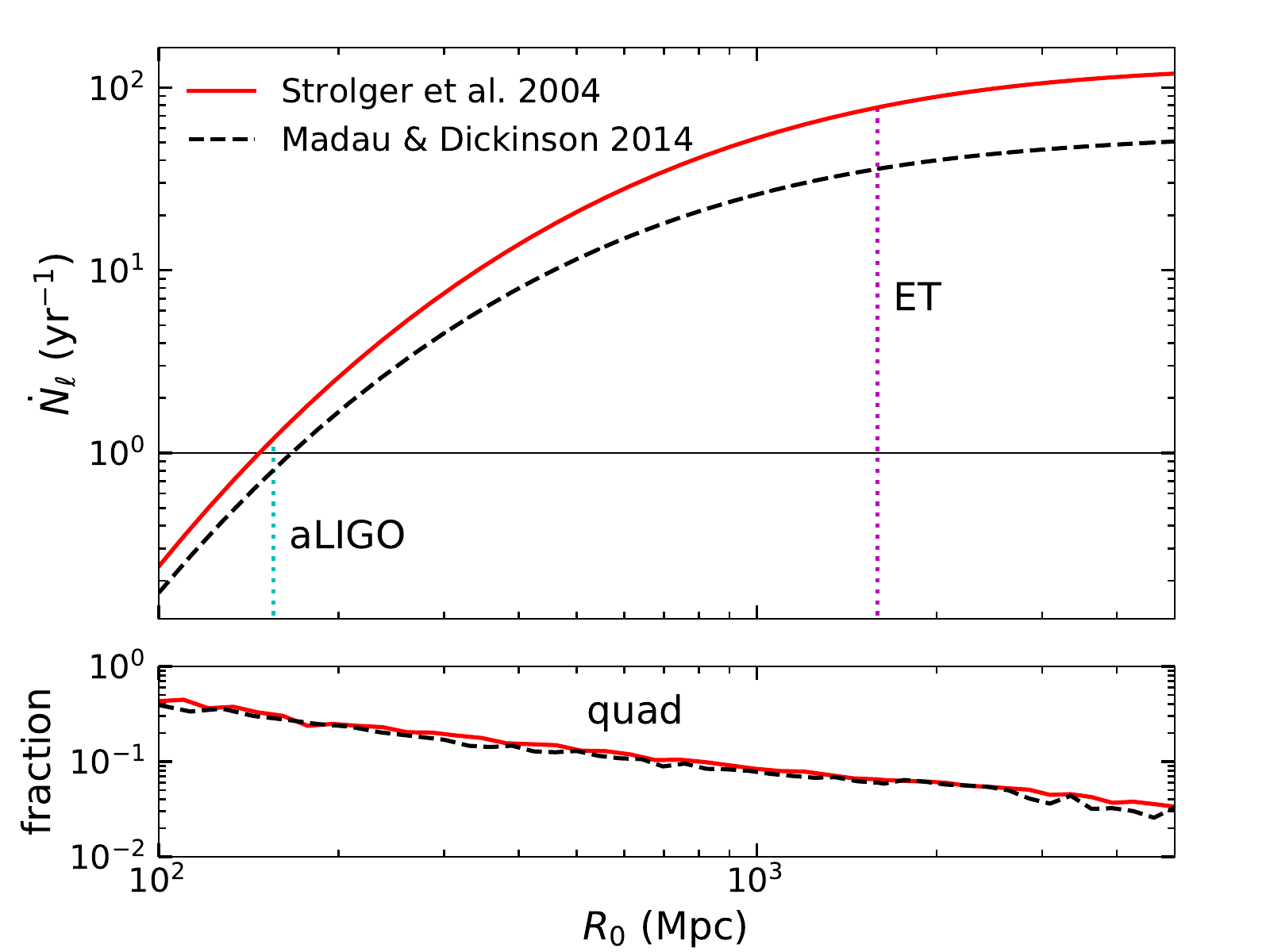}
    \caption{The lensed event rate (top panel) and the fraction of quadruple lenses (bottom panel) as a function of characteristic distance. The red solid and black dashed lines represent results obtained by using the SFR functions from \citet{Strolger2004} and from \citet{Madau2014}, respectively. The vertical cyan and magenta lines show $R_0$ corresponding to aLIGO and ET, respectively. The horizontal line sets the threshold for expectable rate ($1$ event per year). The threshold of S/N is set to be eight.}
    \label{fig:R0}
\end{figure}

We demonstrate the rate of lensed GW events as a function of the characteristic distance in Fig.~\ref{fig:R0}. The notation for the different lines is the same as above. As expected, the larger $R_0$, the larger number of lensed events a detector can observe. This result indicates that any detectors more sensitive to aLIGO are expected to observe several strongly lensed events per year. The declining tendency of the quadruple fraction in the bottom panel is again due to the decrease in the magnification bias.

%
\begin{figure}
	\includegraphics[width=\columnwidth]{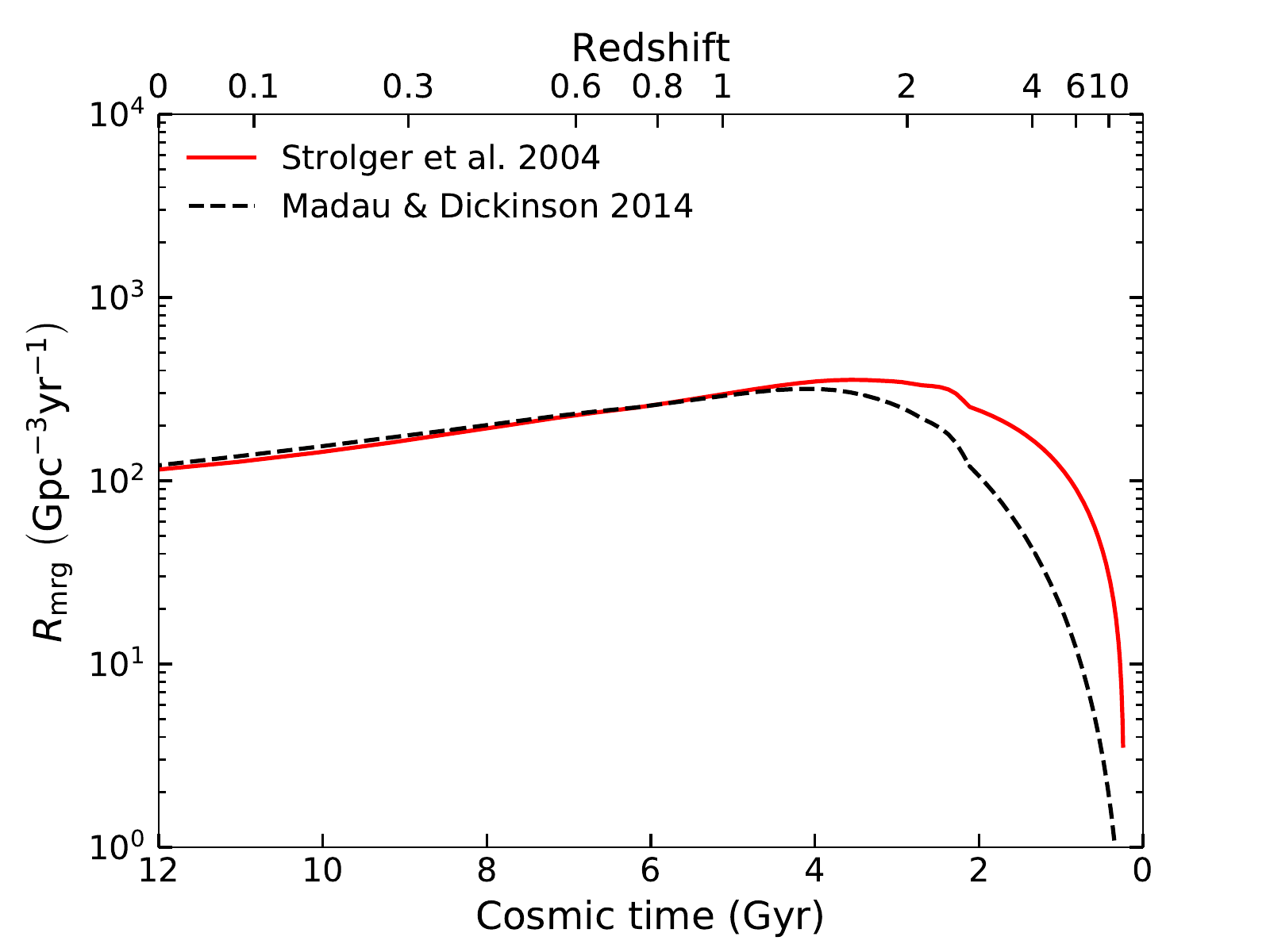}
    \caption{Merger rate density of stellar binary black holes as a function of cosmic time (redshift). The red solid and black dashed lines represent results obtained by using the SFR function from \citet{Strolger2004} and from \citet{Madau2014}, respectively.}
    \label{fig:intrinsic}
\end{figure}

\begin{table*}

	\caption{Prediction for the unlensed and lensed GW event rates in various surveys. We use two different characteristic distance $R_0$ to identify the detectors ($155.4~\text{Mpc for aLIGO};~1591~\text{Mpc for ET}$) and two different SFR functions to estimate the GW source rate (\citealt{Strolger2004} and \citealt{Madau2014}). We adopt the threshold $\rho_0=8$ for all surveys.}
	\label{tab:lensed}
	\centering
	\begin{tabular}{ccccc}
		\hline
        \hline
		\multirow{2}{*}{Detectors} & \multirow{2}{*}{SFR functions} & $\dot{N}_{\text{s}}(>\rho_0)$ & $\dot{N}_{\ell}(>\rho _0)$ & Fraction \\
& & $[\text{yr}^{-1}]$ & $[\text{yr}^{-1}]$ & (quad) \\
		\hline
		\multirow{2}{*}{aLIGO}  & Strolger et al. 2004 & $5.4\times 10^{3}$ & $1.20$ & $0.30$\\
& Madau \& Dickinson 2014 & $5.1\times 10^{3}$ & $0.84$ & $0.26$\\
\hline
		\multirow{2}{*}{ET} & Strolger et al. 2004 & $1.4\times 10^5$ & $79.4$ & $0.06$\\
& Madau \& Dickinson 2014 & $9.6\times 10^4$ & $38.6$ & $0.06$\\
		\hline
	\end{tabular}
\end{table*}

\subsection{Distribution of time delays}
\label{res:time}

A prediction for the time delay distribution is required in order to assess the detectability of multiple images during a finite duration GW survey. We achieve this goal through a semi-analytic technique based on Monte Carlo sampling (see \citealt{Mao1992} for a similar calculation for gamma-ray bursts).

The specific procedure is as follows. First, we randomly generate a sample of $10^7$ lens systems at a given source redshift. The lens objects are considered to be uniformly distributed on the sky, and the lens properties are distributed as described in Section~\ref{lenss}. Then, we solve each lens system to see if it has multiple images, and for those with multiple images, we calculate their time delays through equation (\ref{tau}). By grouping these lens systems according to their time delays, we obtain the distribution of time delays. Since we do not set a threshold of S/N in this calculation, the distribution derived here considers all the lens systems satisfying the lens properties described in Section~\ref{lenss}, not just those observable by a particular survey.

\begin{figure*}
	\includegraphics[width=\textwidth]{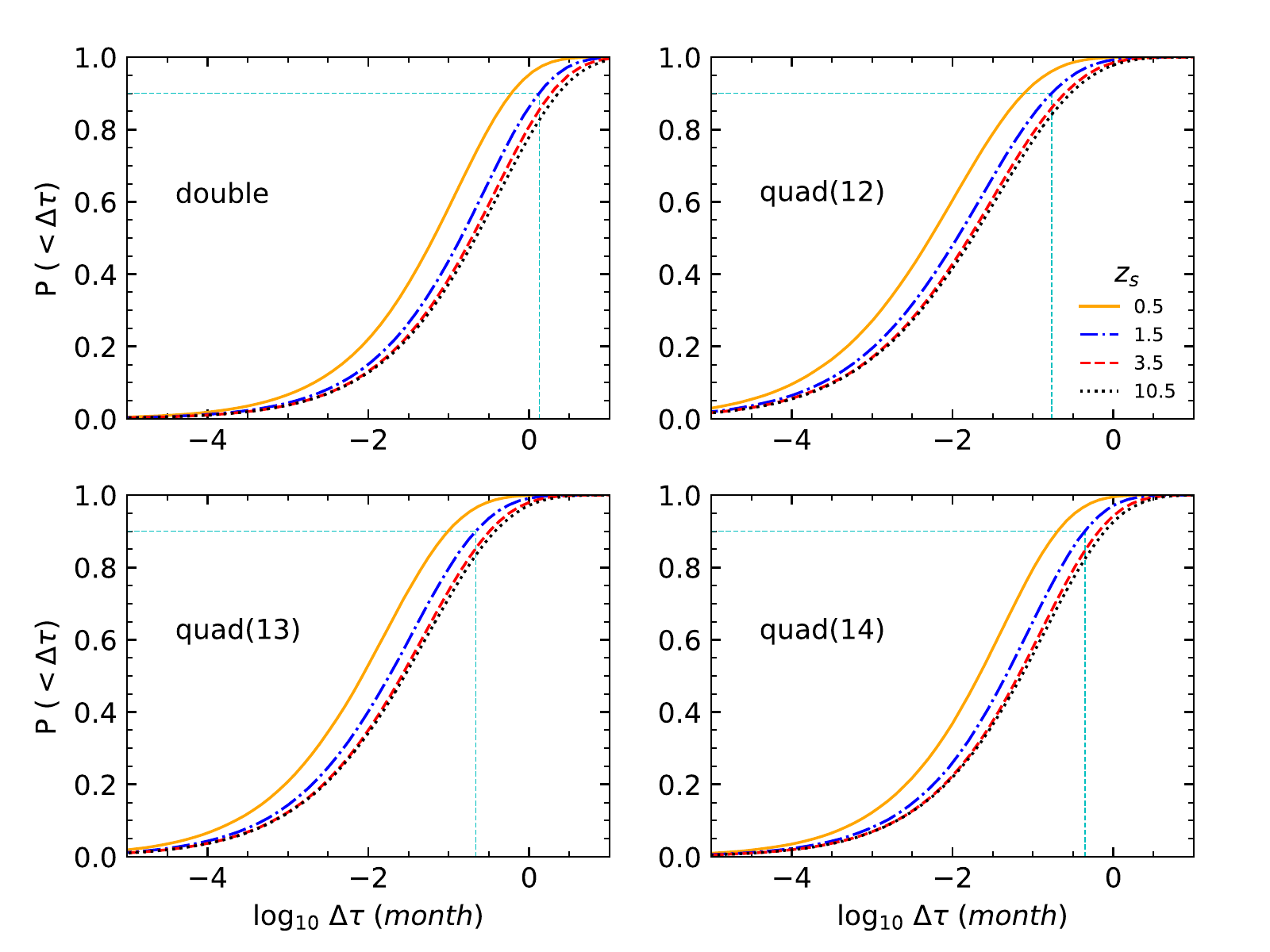}
    \caption{The cumulative distribution function of time delay for various image pairs. From the left- to right-hand panels and top to bottom panels, the four panels show the results for the time delay between the first and the second images for double lenses; and the time delays between the first and the second images [quad(12)], the first and the third images [quad(13)], the first and the fourth images [quad(14)] for quadruple lenses, respectively. Here, the images are named in order of arrival, i.e. 1 indicates the first arrival, 2 is the second arrival, etc. In each panel, four representative source redshifts are shown, i.e. $z_s=$ $0.5$ (orange solid line), $1.5$ (blue dash-dotted line), $3.5$ (red dashed line), and $10.5$ (black dotted line). The cyan dashed line indicates the $90$ per cent cumulative probability for GW sources with $z_{\rm s} =1.5$.}
    \label{fig:dt}
\end{figure*}

Fig.~\ref{fig:dt} shows the cumulative distribution function of the time delay for four representative source redshifts, $z_s=0.5$, $1.5$, $3.5$, and $10.5$, respectively. For double lenses (top left-hand panel) with a typical source redshift ($z_s=1.5$), $90$ per cent of the systems have time delays less than $\sim 1$ month. Even for the systems with a high source redshift ($z_s =10.5$), nearly $80$ per cent have time delays less than $1$ month. Almost all the systems have time delays less than $10$ months. This result indicates that the selection bias raised by the lensing time delay is insignificant, since the data-taking phases of GW detectors in the future will have durations well beyond most lens systems' time delays.\footnote{For example, the first and second runs (O1, O2) of aLIGO lasted for approximately $4$ months and $9$ months, respectively.}

For quadruple lenses, we calculate time delays for three independent image pairs, in order of the arrival time: between the first and the second images [top right-hand panel; quad(12)], between the first and the third images [bottom left-hand panel; quad(13)], and between the first and the fourth images [bottom right-hand panel; quad(14)]. The result shows that for a typical time delay ($z_s=1.5$), $90$ percent of the systems have time delays between the first and the last images shorter than $\sim 0.4$ month, which implies missing any images due to the finite observation duration is unlikely. It is worth pointing out that the time delays between image pairs in quadruple lenses are typically shorter than those in double lenses. This feature implies a possible bias with quadruple lenses being over-represented in a finite survey.

\section{Summary and discussion}
\label{sec:summary}
In this paper, we have investigated the statistical properties of the strong gravitational lensing of GWs from stellar binary black hole coalescences in the context of ground-based detectors. By taking more realistic lens and source properties into account, we make a prediction for the rate of lensed GW events. Moreover, we calculate the probability distribution of lensing time delays to assess the selection bias due to the finite duration of a survey. Our main results can be summarized as follows.

We predict that aLIGO operating at its design sensitivity is expected to detect several lensed GW events (approximately $1$ event per year). The ET prediction is much higher (approximately $40\sim 80$ events per year) due to its much-higher sensitivity. The results are dominated by double lenses, with an expected quadruple fraction of $\sim 30$ per cent for aLIGO events and $\sim6$ per cent for ET events. According to the SFR function from \citet{Madau2014}, the most probable redshifts of the lensed GW sources range from $\sim1.1$ to $\sim2.7$ for aLIGO events and from $\sim1.5$ to $\sim3.7$ for ET events. We emphasize the strong dependence between the predicted lensing rate and the source rate. This dependence leaves space for further improvement of the lensing rate prediction.

Specifically, the estimate of the merger rate density is calibrated to the current observations of stellar binary black hole GW sources by aLIGO and VIRGO, i.e. a mean rate density of $\sim 103~{\rm Gpc^{-3}\, yr^{-1}}$ in the local Universe. However, the current constraint on this mean rate density has a large uncertainty as shown in \citet{Abbott2017}, and it could range from $40$ to $213~{\rm Gyr^{-1}\,yr^{-1}}$ assuming a power-law distribution for the primary black hole masses. Considering this uncertainty, the strongly lensed GW event rate should be in the range of about a factor of $0.4$ to $2.1$ of the estimates listed above (a factor of $\sim5$ uncertainty). Note also that the merger rate density estimated from the simple model presented in this paper seems to be smaller than some estimates by using binary population synthesis models \citep[see discussion in][]{CaoL2017}. This may suggest that the strongly lensed GW event rate, especially for ET, may be even larger than the estimates obtained here.

Furthermore, we have developed a general calculation formalism of the lensing rate, not restricted to any specific detectors. The result indicates that any ground-based detectors more sensitive than aLIGO are anticipated to observe several lensed GW events per year (see Fig.~\ref{fig:R0}). Detectors need not be a single instrument with unprecedented sensitivity such as ET but can be a network of interferometers such as aLIGO together with Virgo. As networks of GW detectors become routine in the near future (e.g. \citealt{Abbott2016LRR}, for a review of the commissioning roadmap), the detection of lensed GW events is expected even before ET becomes operational.

We have evaluated the chance of missing some images in a finite observation period, by examining the probability distribution of the lensing time delays. We find most lens systems involved in this study have time delays less than $\sim1$ month (see Fig.~\ref{fig:dt}). Since GW surveys in the future will have a duration much longer than a month, we expect the selection bias raised by the finite observation time should be small. Nevertheless, we emphasize that the time delays of quadruple lenses are systematically smaller than those of double lenses due to the smaller impact parameters in quadruple systems (the source is closer to the centre of the lens galaxy). This feature may result in a slightly higher fraction of quadruple lenses in a finite observation period.

In a real GW survey, other factors besides the finite duration may cause the absence of some images from detection, such as unexpected glitches in the detector, detector downtime for improvement and so on. Most of these effects can be eliminated by building up a network of several detectors (see \citealt{Abbott2017c}, for a treatment of the glitch in a real GW observation). This implies another advantage of joint detection in GW astronomy.

There are also some systematic errors due to the uncertainty of the velocity distribution function of lensing galaxies (equation~\ref{equ:VDF}). In this work, we adopt the modified Schechter function with parameters from \citet{Choi2007} based on the SDSS DR3 data, while other authors using different data bases obtain somewhat different parameters (see \citealt{Montero2017}, for a recent comparison). Also, different strategies for sample selection and function modelling can affect the shape of the velocity distribution function (see e.g. \citealt{Sohn2017}). Furthermore, the velocity distribution function is expected to evolve with time at high redshift, though the details of this evolution are somewhat uncertain (see e.g. \citealt{Bezanson2011}). All these factors may introduce uncertainties to the results.

In this paper, we only consider GWs arising from stellar binary black hole coalescences. Although these sources as a whole dominate the high-band GW events, there are other types of double compact objects that can generate GWs, such as the inspiral of neutron star-neutron star or black hole-neutron star binaries. These sources are especially intriguing in multimessenger observations, as these systems are believed to be associated with kilonovae and can produce electromagnetic counterparts (see e.g. \citealt{Metzger2012}, for a theoretical study and \citealt{Abbott2017ApJ} for a real observation). In consideration of these systems' enormous potential for physical and cosmological research (see e.g. \citealt{Collett2017,Wei2017}), a further statistical study involving these systems is warranted.

\section*{Acknowledgements}

We thank Richard Long for many constructive comments that improved the paper. This work was supported by the National Natural Science Foundation of China (Grant No. 11333003, 11390372 to SM; and 11690024 and 11390372 to YL). YL was also partly supported by the Strategic Priority Program of the Chinese Academy of Sciences (Grant No. XDB 23040100), and the National Key Program for Science and Technology Research and Development (Grant No. 2016YFA0400704).



\bibliographystyle{mnras}
\bibliography{bibfile} 



\appendix

\section{Lens Theory}
\label{appGL}

In this appendix, we present further details of the lens theory based on the SIE model with external shear which is used in this paper. We refer the interested reader to \cite{Schneider1992}; \cite{Schramm1990};   \cite{Kochanek1991}; \cite{Keeton1998}, and references therein for thorough discussions.

Bearing in mind the two-dimensional nature of the lensing calculation, we adopt $(x,y)$ and $(x_s,y_s)$ as position vectors in the lens plane (the ``thin lens'' approximation) and source plane, respectively. Using equations (\ref{SIE}) and (\ref{shear}), the first derivatives of the lens potential $\phi(=\phi^{\rm SIE}+\phi^{\rm shear})$ are shown as (e.g. \citealt{Keeton1998})
\begin{equation}
\begin{split}
\phi_x=&\frac{b_I(q)}{e}\text{arctanh}\left[ \frac{ex}{\psi}\right]+x\gamma_1+y\gamma_2~,\\
\phi_y=&\frac{b_I(q)}{e}\arctan\left[\frac{ey}{\psi}\right]-y\gamma_1+x\gamma_2~,
\end{split}
\end{equation}
where $b_I(q)=\lambda(q)\sqrt{q}$, $\psi=\sqrt{x^2+q^2y^2}$, and $e=\sqrt{1-q^2}$ is the eccentricity of lensing galaxies. The second derivatives are
\begin{equation}
\begin{split}
\phi_{xx}=&\frac{b_I(q)}{\psi}\frac{y^2}{x^2+y^2}+\gamma_1~,\\
\phi_{yy}=&\frac{b_I(q)}{\psi}\frac{x^2}{x^2+y^2}-\gamma_1~,\\
\phi_{xy}=&-\frac{b_I(q)}{\psi}\frac{xy}{x^2+y^2}+\gamma_2~.
\end{split}
\end{equation}

Generally, the lens equation
\begin{equation}\label{lens}
\begin{split}
x_s=& x-\phi_x(x,y)\\
y_s=& y-\phi_y(x,y)
\end{split}
\end{equation}
is a system of nonlinear equations. Directly employing numerical calculation to solve equation (\ref{lens}) could be time-consuming. Introducing the polar coordinates $x=r\cos\alpha$, $y=r\sin\alpha$, (\ref{lens}) becomes
\begin{equation}\label{lens21}
\begin{split}
&\left(y_s+\phi_y^{\rm SIE}\right)\left[(1-\gamma_1)\cos\alpha-\gamma_2\sin\alpha\right]\\
&=\left(x_s+\phi_x^{\rm SIE}\right)\left[(1+\gamma_1)\sin\alpha-\gamma_2\cos\alpha\right]~,
\end{split}
\end{equation}
and
\begin{equation}\label{lens22}
\begin{split}
r &=\left(y_s+\phi_y^{\rm SIE}\right)/\left[(1+\gamma_1)\sin\alpha-\gamma_2\cos\alpha\right]\\
&=\left(x_s+\phi_x^{\rm SIE}\right)/\left[(1-\gamma_1)\cos\alpha-\gamma_2\sin\alpha\right]~,
\end{split}
\end{equation}
with
\begin{equation}
\begin{split}
\phi_x^{\rm SIE}=&\frac{b_I(q)}{e}\text{arctanh}\left[\frac{e\cos \alpha}{\sqrt{\cos(\alpha)^2+q^2\sin(\alpha)^2}}\right]~,\\
\phi_y^{\rm SIE}=&\frac{b_I(q)}{e}\arctan\left[\frac{e\sin \alpha}{\sqrt{\cos(\alpha)^2+q^2\sin(\alpha)^2}}\right]~.
\end{split}
\end{equation}
Now we can numerically solve the one-dimensional equation (\ref{lens21}) to obtain the polar angle $\alpha$, then substitute $\alpha$ into equation (\ref{lens22}) to obtain the radius $r$.

The time delay $\tau$ and magnification $\mu$ are given by (e.g. \citealt{Schneider1992})
\begin{equation}\label{tau}
\tau = (1+z_{\ell})\frac{D_{\ell}D_s}{D_{ls}}\theta_E^2\left\{\frac{1}{2}\left[(x-x_s)^2+(y-y_s)^2\right]-\phi(x,y)\right\}~,
\end{equation}
and
\begin{equation}\label{magsiee}
\mu = (1-\phi_{xx}-\phi_{yy}-\phi_{xy}^2+\phi_{xx}\phi_{yy})^{-1}~,
\end{equation}
where $z_{\ell}$ is the redshift of the lens, and $D_{\ell}$, $D_s$, $D_{\ell s}$ denote the angular distances to the lens, the source and between the lens and the source, respectively. The angular Einstein radius is given by $\theta_E \equiv 4{\rm\pi}\sigma_v^2(D_{\ell s}/D_{s})$.

\section{Binary Black Hole Merger Rate}
\label{source}

In this appendix, we describe the approach we use in computing the merger rate of binary black holes. We consider only stellar binary black boles formed from isolated massive binary stars in galaxies. These are the most promising sources of GWs that can be detected by ground-based GW surveys. The approach is similar to that presented in \citet{CaoL2017} and \citet{Dvorkin2016}.

Generally, the birth rate per unit volume of single black holes with mass $M_{\bullet}$ at the cosmic time $t$ is given by
\begin{equation}
\begin{split}
R_{\rm birth}(M_\bullet;t) =& \int dm_\star~\phi(m_\star)\\
&\times \int  dZ~\dot{\psi}(Z;t)~ \delta \left[m_\star - g^{-1}(M_{\bullet},Z)\right]~.
\end{split}
\end{equation}
Here $\phi(m_\star)$ is the initial mass function of the star with the Chabrier initial mass function \citep{Chabrier2003} being adopted, $\dot{\psi}(Z;t)$ is the star formation rate (SFR) per unit volume with metallicity $Z$ at cosmic time $t$, and $\delta$ is Dirac-$\delta$ function. The relation between the mass of a stellar remnant black hole and the mass of its progenitor star is given by $M_{\bullet} = g(m_\star,Z)$. We adopt the version obtained by \citet{Spera2015}.

We assume that $\dot{\psi}(Z;t)$ can be separated into two independent
functions, one is the total SFR function at redshift $z$ and the other is the metallicity
distribution function at that redshift. For the total SFR function, we adopt the observationally determined functions from \citet{Madau2014} and from \citet{Strolger2004}. For the metallicity distribution function, we adopt the mean metallicity given by \citet{Belczynski2016}.

Assuming that a fraction ($f_{\rm eff}$) of black holes exist as the primary components\footnote{The primary component of a binary has mass ($M_{\bullet,1}$) larger than that ($M_{\bullet,2}$) of the secondary one.} of binaries which can merge within the \emph{Hubble} time, the merger rate density of stellar binary black holes is then given by
\begin{equation}
R_{\rm mrg}(M_{\bullet,1},q;z) =f_{\rm eff}\int dt_{\rm d}~ R_{\rm
birth}(M_{\bullet,1};z)~ P_t(t_{\rm d})~ P_q(q)~.
\end{equation}
Here $P_t(t_{\rm d})$ is the probability distribution of the time delays $t_{\rm d}$ between the formation of stellar binary black holes and merger. We adopt the form $P_t (t_{\rm d}) \propto t_{\rm d}^{-1}$ \citep{OShaughnessy2010, Belczynski2016, Lamberts2016} and assume the minimum and maximum values of $t_{\rm d}$ are $50$ Myr and the \emph{Hubble} time, respectively. $P_q(q)$ is the probability distribution of the mass ratio $q=M_{\bullet,2}/M_{\bullet,1}$ and is assumed to be independent of the black hole mass. We assume $P_q(q) \propto q$ over the range from $0.5$ to $1$, which seems to be consistent with binary population synthesis results \citep{Belczynski2016,CaoL2017}. The parameter $f_{\rm eff}$ is determined by adopting the constraint on
the mean detection rate of $103~ {\rm Gpc}^{-3}\,{\rm yr}^{-1}$ given by the current aLIGO detection \citep{Abbott2017} to calibrate the merger rate density at $z\sim 0$ obtained from the model.

The merger rate density with respect to chirp mass $\mathcal{M}_0$ at redshift $z$ can be obtained as
\begin{equation}
\label{equ:merger}
\begin{split}
R_{\rm mrg}(\mathcal{M}_0;z) = \iint & dM_{\bullet,1}~ dq~R_{\rm mrg}(M_{\bullet,1},q;z)\\
 &\times \delta(\mathcal{M}_0 - \mathcal{M}_{q,M_{\bullet,1}})~,
\end{split}
\end{equation}
where $ \mathcal{M}_{q,M_{\bullet,1}}=q^{3/5}M_{\bullet,1}/(1+q)^{1/5}$ is the chirp mass of a black hole  binary with primary black hole mass $M_{\bullet,1}$ and the mass ratio $q$.

By marginalizing over $\mathcal{M}_0$ in equation~(\ref{equ:merger}), we can obtain the merger rate density at redshift $z$:
\begin{equation}\label{equ:merger2}
  R_{\rm mrg}(z) = \int d\mathcal{M}_0~R_{\rm mrg}(\mathcal{M}_0;z)~.
\end{equation}


\bsp	
\label{lastpage}
\end{document}